\documentstyle[prl,aps,12pt]{revtex}

\def\be{\begin{equation}}
\def\ee{\end{equation}}
\def\bea{\begin{eqnarray}}
\def\eea{\end{eqnarray}}

\input epsf.sty
\begin{document}
\draft
\title{\large\bf Counterterms/M-Theory Corrections
to D=11  Supergravity  }
\bigskip
\author{S. Deser and 
        D. Seminara\footnote{\scriptsize Currently: Laboratoire de Physique
        Th\'eorique de L' \'Ecole Normale Sup\'erieure,
         24 Rue Lhomond-75231 Paris CEDEX 05}}
\address{\it Department of  Physics, Brandeis University,
                        Waltham, MA 02454, USA\\
}
\bigskip
\date{Received \today}
\maketitle
\bigskip
\font\ninerm = cmr9
\pagestyle{empty}
\begin{abstract}
\ninerm\noindent We construct a local  on-shell  invariant in
D=11  supergravity from the nonlocal four-point tree scattering amplitude.
Its existence, together 
with earlier arguments, implies  non-renormalizability of 
the theory at lowest possible, two loop, level. This invariant, 
whose leading  bosonic terms are exhibited,
may also express  the leading, ``zero-slope", M$-$theory 
corrections to its D=11 supergravity limit.
\end{abstract}
\pacs{PACS numbers:\ \ }
\vfill
\begin{flushright}
BRX-TH-443
\end{flushright}
\newpage
\pagestyle{plain}
\medskip
With the advent of $M-$physics, of which it is the local limit, D=11
supergravity \cite{CrJuSc} has regained a central role. This
connection adds a further motivation to our quest
for its explicit on-shell supersymmetric (SUSY) invariants: Not
only would their  existence describe specific candidate counterterms,
completing a recent argument for the field theory's 
nonrenormalizability, but they 
would also exemplify concrete ``zero-slope'' corrections from the full
$M-$theory (whatever its ultimate form) similar to 
the corresponding string corrections to their limiting, $D=10$,
supergravities. That such invariants had not been given earlier
is due to the absence  of a systematic supersymmetric calculus, 
or even of a practical way to verify candidate  terms. Indeed, it was 
only very  recently  \cite{Green,Russo} that the tensorial
structure  of the relevant invariant's four graviton sector  was found, by
explicit one-loop calculations.
The present effort originated in trying to
generalize techniques known from lower-dimensional models, such as 
the use of gravitational Bel-Robinson (BR) tensors as currents in 
constructing D=4 supergravity invariants \cite{DKS}; despite the strong
degeneracy  in the number of such currents at D=4, their 
extensions will  indeed
play a key part in our  D=11 construction. A major step
forward  in this area was recently made \cite{BDetal} through beautiful
use of the YM SUSY/SUGRA open/closed string correspondence,
analytically extended to its maximal (D=10) dimension. Although there
is no underlying D=11 YM SUSY model, we shall argue that 
the construction of  \cite{BDetal}
together with the invariant provided here, lend 
strong credence to  a 2-loop nonrenormalizability verdict 
for D=11 supergravity.

Our construction is a physical one, with manifest  supersymmetry: we
calculate the tree-level four-point scattering amplitudes of the 
D=11 theory.
This
procedure  has several merits: First, there is no fermion-boson mixing
in the tree diagrams; hence we are free just to calculate 
the bosonic contributions knowing that they are part of a guaranteed SUSY
invariant, namely the total four-point S-matrix. Second, because SUSY 
transformations are linear to leading 
order, there is no mixing with higher-point amplitudes. 
Third, we will see that one can uniformly extract the desired, 
local, invariant from the nonlocal $S-$matrix without loss of SUSY. 
The basis for  
our computations is the  full action of \cite{CrJuSc}, expanded to
the order required for obtaining the four-point scattering amplitudes
among its two bosons, namely the graviton and the three-form potential
$A_{\mu\nu\alpha}$ with field strength $F_{\mu\nu\alpha\beta}\equiv
4\partial_{[ \mu} A_{\nu\alpha\beta]}$, invariant under the gauge
transformations
$\delta A_{\mu\nu\alpha}=\partial_{[\mu}\xi_{\nu\alpha ]}$. From the 
bosonic truncation of this action (omitting obvious summation indices),
\begin{equation}
\label{Lagra}
I^B_{11}=\int d^{11}x \left [-\frac{\sqrt{g}}{4\kappa^2} R(g)
-\frac{\sqrt{g}}{48} F^2
+\frac{2\kappa}{144^2}\epsilon^{1\cdots 11}
F_{1\cdots}F_{5\cdots}A_{..11}\right],
\end{equation}
we extract the relevant vertices and propagators; note that
$\kappa^2$  has dimension $[L]^9$ and that the (P, T) conserving
cubic Chern-Simons (CS)  term
depends explicitly on $\kappa$ but is (of course) gravity-independent.
The propagators come from the quadratic terms in
$\kappa h_{\mu\nu}\equiv g_{\mu\nu}-\eta_{\mu\nu}$ and $A_{\mu\nu\alpha}$;
they need no introduction. There are three cubic vertices, namely
graviton, pure form and mixed form-graviton that we schematically
represent as
\begin{mathletters}
\label{gggl}
\begin{eqnarray}
\label{ggg}
&&V_3^g \sim (\partial h\partial h) h \equiv
 \kappa T^{\mu\nu}_g h_{\mu\nu},
\ \ \ V_3^{gFF}\equiv \kappa T^{\mu\nu}_F h_{\mu\nu},\ \ \
V_3^F\sim \kappa 
\epsilon A  F F \equiv \kappa
A_{\mu\nu\alpha}C^{\mu\nu\alpha}_F\\
\label{ggg1}
&&T_g^{\mu\nu}\equiv G_{(2)}^{\mu\nu} , \ \ 
T^{\mu\nu}_F\equiv 
F^{\mu}F^{\nu}-\displaystyle{\frac{1}{8}}\eta^{\mu\nu} F^2,\ \ 
C_F^{\rho\sigma\tau}\equiv\frac{2}{(12)^4}
\epsilon^{\rho\sigma\tau\mu_1\cdots\mu_8}
F_{\mu_1\cdots} 
F_{\cdots\mu_8}.
\end{eqnarray}
\end{mathletters}
The form's current $C_F$ and stress tensor $T_F$ 
are both manifestly gauge 
invariant. In our computation, two legs of the 
three-graviton  vertex are always on linearized  
Einstein shell; we 
have exploited this fact in writing it in the simplified form 
(\ref{gggl}), the subscript on the Einstein tensor
denoting its quadratic part in $h$. 
[Essentially, the on-shell legs are the ones in
$T_g^{\mu\nu}$, the off-shell one  multiplies it.]
To achieve  coordinate 
invariance to correct, quadratic, order one must also include 
the four-point contact vertices
\begin{equation}
V^{g}_4\sim \kappa^2 (\partial h\partial h) h h, \ \ \ V_4^{gF}
=\kappa^2  \frac{\delta I^F}{\delta g_{\alpha\beta}\delta g_{\mu\nu}}
h_{\alpha\beta} h_{\mu\nu}
\end{equation}
when calculating the amplitudes; these are the remedies for the
unavoidable coordinate variance of the gravitational stress
tensor $T^{\mu\nu}_g$ and the fact that  $T^{\mu\nu}_F h_{\mu\nu}$ is 
only first order coordinate-invariant. The  
gravitational vertices are not given
explicitly, as they are both horrible 
and well-known \cite{Sannah,VandeVen}. We  reiterate that gravitinos 
are decoupled  at tree level; while four-point amplitudes  involving 
them would mix with  bosonic ones under supersymmetry transformations,
this  would merely provide a (useful!) check on our arithmetic. 

We start with the $4-$graviton amplitude, obtained by contracting two
$V^g_3$ vertices in all three  channels (labelled by the Mandelstam 
variables $(s,t,u)$) through an
intermediate  graviton propagator (that provides a single
denominator);  adding  the contact $V^g_4$
and then setting the external
graviton polarization tensors on free Einstein shell.
The resulting amplitude $M^g_4 (h)$ will be a nonlocal (precisely
thanks to the local $V^g_4$ contribution!) quartic in the
Weyl tensor\footnote{ We do not differentiate in
notation between Weyl and Riemann here and  also express
amplitudes in covariant terms for simplicity, even though  they are
only valid to   lowest relevant order in the linearized
curvatures.}.
Explicit calculation is of course required to discover the exact $R^4$
combinations involved and things are much more complicated in higher
dimensions than in D=4, where there are exactly two possible local 
quartics in the Weyl tensor, for example the squares of Euler 
($E_4\equiv R^* R^*$) and Pontryagin ($P_4 \equiv R^* R$) densities  
($^\star \!R \equiv 1/2 \epsilon R$). The special property of the Einstein
action (that also 
ensures its supersymmetrizability) is that this amplitude must be 
maximally helicity conserving  (treating all particle as incoming), 
thereby  fixing its local part to be \cite{DGetal}
$(E_4-P_4)(E_4+P_4)$. This invariant is also, owing to identities
peculiar to D=4,
expressible \cite{DKS} as the square of the (unique in $D=4$) BR tensor 
$B_{\mu\nu\alpha\beta}=(R R +R^* R^*)_{\mu\nu\alpha\beta}$. But D=4
is a highly degenerate case in both  respects: generically,
there are seven independent quartic monomials \cite{fulling} 
in the Weyl tensor for ${\rm D}\ge8$ and an intrinsically three-parameter 
family of $BR$ tensors; as might be expected, there is no longer any 
simple  equivalence between $(BR)^2$ forms and helicity (though it 
might be fruitful to explore its extensions to generic D). Still, 
these descriptions are robust: for example, one hint for the 
gravitational amplitude is provided  by its diagrammatic origin in
terms of $T^{g}_{\mu\nu}$ because there is a (highly 
gauge-dependent) identity  of the 
schematic form $B_{\mu\nu\alpha\beta}\sim 
\partial^2_{\alpha\beta}T_{\mu\nu}^g$. Within  our space limitations, 
we cannot exhibit  the actual calculation here; fortunately, this 
amplitude  has 
already  been given  (for  arbitrary $D$) in the pure gravity
context \cite{Sannah}. It can be shown, using the
basis of \cite{fulling}, to be of the form
\begin{equation}
\label{pippo}
M^{g}_4=\kappa^2 (4 s t u)^{-1} 
t_8^{\mu_1\cdots\mu_8} t_8^{\nu_1\cdots\nu_8}
R_{\mu_1\mu_2\nu_1\nu_2}R_{\mu_3\mu_4\nu_3\nu_4}R_{\mu_5\mu_6\nu_5\nu_6}
R_{\mu_7\mu_8\nu_7\nu_8}\equiv  (s t u)^{-1} L^g_4,
\end{equation}
up to a possible contribution from the quartic Euler density
$E_8$, which is  a total divergence to this order   
(if present, it would only contribute at $R^5$ level). The 
result (\ref{pippo})
is also the familiar superstring zero-slope limit
correction to D=10 supergravity, where the
$t_8^{\mu_1\cdots\mu_8}$ symbol originates from the
D=8 transverse subspace\cite{Schwartz}. [Indeed, the ``true'' origin
of the ten
dimensional analog of (\ref{pippo}) was actually traced back to 
D=11 in the one-loop computation of \cite{Green,Russo}.] 
Note that the local part,
$L^g_4$, is simply extracted through 
multiplication of $M^g_4$ by $s t u $, which in no way alters SUSY
invariance, because all parts of $M_4$ behave the same way.  

In many respects, the form (\ref{pippo}) for the 
$4-$graviton contribution is a perfectly physical one. However in
terms of the rest of the invariant to be obtained below, one would like a
natural formulation with currents that encompass both gravity and
matter in a unified way as in fact occurs in e.g. $N=2$, D=4
supergravity \cite{DK}. This might also lead to some understanding of
other SUSY multiplets.  Using the
quartic basis expansion, one may  rewrite  $L^g_4$ in various ways
involving conserved $BR$ currents and a closed $4-$form
$P_{\alpha\beta\mu\nu}= 1/4 R^{ab}_{[\mu\nu} R_{\alpha\beta]a b}$, for
example 
\begin{mathletters}
\label{lippi}
\begin{eqnarray} 
&&L^g_4=48 \kappa^2\left[ 2 B_{\mu\nu\alpha\beta}B^{\mu\alpha\nu\beta}
-B_{\mu\nu\alpha\beta}B^{\mu\nu\alpha\beta}+
P_{\mu\nu\alpha\beta}P^{\mu\nu\alpha\beta}+6
B_{\mu\rho\alpha}^{\ \ \ \ \rho}B^{\mu\sigma\alpha}_{\ \ \ \ \sigma}
-\frac{15}{49} (B^{\mu\nu}_{\ \ ~\mu\nu})^2\right],\\
&&B_{\mu\nu\alpha\beta}\equiv
R_{(\underline{\mu}\rho\alpha\sigma} 
R^{\ ~\rho\ ~\sigma}_{\underline{\nu})\ ~\beta\ }
-\frac{1}{2}g_{\mu\nu}
R_{\alpha\rho\sigma\tau} R_{\beta}^{\ ~\rho\sigma\tau}-
\frac{1}{2}g_{\alpha\beta}
R_{\mu\rho\sigma\tau} R_{\nu}^{\ ~\rho\sigma\tau}
+\frac{1}{8} g_{\mu\nu} g_{\alpha\beta}
R_{\lambda\rho\sigma\tau}R^{\lambda\rho\sigma\tau},
\end{eqnarray}
\end{mathletters}
where $(~)$ means symmetrization 
with weight one of the underlined indices. 
At D=4, $P_{\mu\nu\alpha\beta}$ obviously reduces to 
$\epsilon_{\mu\nu\alpha\beta} P_4$, and $L_4^g$ can easily be shown 
to have the correct $B^2$ form, as must be the case from brute force 
dimensional reduction arguments.

Let us now turn to the pure form amplitude, whose
operative currents are the Chern-Simons $C^F_{\mu\nu\alpha}$ and 
the stress tensor $T^F_{\mu\nu}$, mediated respectively by the $A$
and graviton propagators; each contribution is separately
invariant. By dimensions, the building block will be  $\kappa
\partial F\partial F$; getting hints from D=4, however, would require
using (unwieldy!) $N=8$ models.
Instead, we computed the two relevant, 
$C_F C_F$ and $T_F T_F$, diagrams directly, 
resulting in the four-point amplitude\footnote{An independent calculation
of $M^F$ has just been reported in \cite{12}; we have not compared
details.}    
$M^F_4= (s t u)^{-1} L^F_4= (s t u)^{-1} \kappa^2 (\partial F)^4$,
again with an overall ($stu$) factor. 
An economical 
way to organize $L^F_4$ is in terms of matter BR tensors
and corresponding $C^F$ extensions, prototypes
being  the ``double gradients'' of $T^F_{\mu\nu}$
and of $C^F$,
\begin{mathletters} 
\label{BF}
\begin{eqnarray}
&&B^F_{\mu\nu\alpha\beta}=
\partial_{\alpha} F_{\mu}
\partial_{\beta} F_{\nu}+
\partial_{\beta} F_{\mu}
\partial_{\alpha} F_{\nu}-
\frac{1}{4}\eta_{\mu\nu} \partial_{\alpha} F
\partial_{\beta} F, \ \ \ \partial^\mu B^F_{\mu\nu\alpha\beta}=0,\\
&&C^F_{\rho\sigma\tau;\alpha\beta}
=\frac{1}{(24)^2}\epsilon_{\rho\sigma\tau\mu_1\cdots\mu_8}
\partial_\alpha F^{\mu_1\cdots\mu_4} \partial_\beta
F^{\mu_5\cdots\mu_8},\ \ \ \partial^\rho C^F_{\rho\sigma\tau;\alpha\beta}=0.
\end{eqnarray}
\end{mathletters}
From (\ref{BF}) we can construct $L^F_4$ as
\begin{equation}
\label{BF1}
L^F_4=\frac{\kappa^2}{36}B^F_{\mu\nu\alpha\beta}
B^{F}_{\mu_1\nu_1\alpha_1\beta_1} 
G^{\mu\mu_1;\nu_1\nu}K^{\alpha\alpha_1;\beta_1\beta}-\frac{\kappa^2}{12}
C^F_{\mu\nu\rho;\alpha\beta}C^{F\mu\nu\rho}_{~~~~~~\alpha_1\beta_1}
K^{\alpha\alpha_1;\beta_1\beta}.
\end{equation}
The matrix 
$G^{\mu\nu;\alpha\beta}\equiv 
\eta^{\mu\alpha} \eta^{\nu\beta}+\eta^{\nu\alpha} \eta^{\mu\beta}- 
2/(D-2)\eta^{\mu\nu} \eta^{\alpha\beta} $ is the usual numerator 
of the graviton propagator on conserved sources. The origin of
$K^{\mu\nu;\alpha\beta}\equiv \eta^{\mu\alpha} \eta^{\nu\beta}+
\eta^{\nu\alpha} \eta^{\mu\beta}- \eta^{\mu\nu} 
\eta^{\alpha\beta} $ can be traced back to ``spreading" the 
$stu$ derivatives: for example, in the $s-$channel, {\it e.g.}, we can 
write $t u=-1/2  K^{\mu\nu;\alpha\beta}p^1_\mu p^2_\nu
p^3_\alpha p^4_\beta$; the analogous identities for the other 
channels can be obtained by crossing\footnote{It is convenient to define
$s\equiv(p_1\cdot p_2), ~t\equiv(p_1\cdot p_3),~u\equiv(p_1\cdot
p_4)$, with $p_1+p_2=p_3+p_4$. Note also the absence of  $(G,K)$ factors
from (5), they are  already  incorporated into the $BR$'s.}.
It is these identities that enabled us to write 
$M_4$'s universally as $(stu)^{-1} L_4$'s: Originally the $M_4$ have 
a single denominator (from the intermediate specific exchange, $s-$,
$t-$ or $u-$channel); we uniformize  them all to $(stu)^{-1}$ through 
multiplication of say $s^{-1}$ by $(tu)^{-1} (tu)$. The extra
derivatives thereby distributed  in the numerators have the further
virtue of turning all polarization tensors into curvatures and 
derivatives  of forms, as we have indicated. It is worth
noting that the matter $(BR)^2$ form  (\ref{BF1}) is in fact  valid
for any matter-matter four-point amplitude mediated by a graviton
through  minimal coupling, simply because of the  $h_{\mu\nu}
T^{\mu\nu}_{matt.}$ vertex and the $BR_{matt.}\sim\partial^2
T_{matt.}$ relation. In particular, one can easily give natural
extensions  of the bosonic results both for the pure fermionic
$4-$point function, since it 
too has an
associated BR tensor $\sim \partial^2_{\alpha\beta} 
T^{\psi}_{\mu\nu}$ and for mixed
fermi-boson contributions. For example,the former resembles (7), with a 
$B^\psi B^\psi$  part  as well as a $C^\psi C^\psi$
part from the nonminimal 
$\bar{\psi} \Gamma\psi F$ coupling in $I_{11}$. Indeed
``current-current'' terms are generically present  for any 
amplitude generated by any gauge-field-current coupling, 
as evidenced by these ubiquitous  $C C$ contributions.

The remaining amplitudes are the form  ``bremsstrahlung'' $M^{FFFg}$ and
the graviton-form scattering $M^{Fg}_4$.  The $M^{FFFg}$ amplitude 
represents radiation of a graviton from the CS term, {\it i.e.}, 
contraction of the  $CS$ and  $T^F_{\mu\nu}h^{\mu\nu}$ vertices by 
an intermediate $A-$line, yielding 
\begin{mathletters}
\begin{eqnarray}
&&M_4^{FFFg}= (s t u)^{-1} L_4^{FFF g}, \ \ \ L_4^{FFF g}=  
-\frac{\kappa^2}{3} C^F_{\mu\nu\rho;\alpha\beta}
C^{RF\mu\nu\rho}_{~~~~~~~\alpha_1\beta_1}K^{\alpha\alpha_1;\beta_1\beta},\\
&&C^{RF}_{\mu\nu\rho;\alpha\beta}
\equiv
4\partial_\lambda\left( R^{\sigma\  [\lambda
}_{\ (\alpha\ ~\beta)} F_{\sigma}^{\ \mu\nu\rho]}\right )
-\frac{2}{3}R^{~\sigma\  ~\lambda
}_{\ ~(\alpha\ ~\beta)}\partial_\lambda F_{\sigma}^{\ \mu\nu\rho} \; .
\end{eqnarray}
\end{mathletters}
The off-diagonal current $C^{RF}$ has antecedents in $N=2$
D=4 theory \cite{DK}; it is unique only up to terms vanishing 
on contraction with $C^F$. While  its (8b) form is compact, there
are more promising variants, with better conservation and trace properties.
The $M^{Fg}, \sim\kappa^2  R^2 
(\partial F)^2$, has three distinct diagrams: mixed $T^F
T^g$ mediated by the graviton; gravitational Compton amplitudes  
$\sim(hh)T_F T_F$ with a virtual $A-$line, and finally the $4-$point 
contact vertex $F F h h$. The resulting  $M_4^{Fg}$ is again
proportional  to $(s t u)^{-1}$, 
\begin{equation}
M^{Fg}_4= (s t u)^{-1} L_4^{Fg}, \ L_4^{Fg}
=\frac{\kappa^2}{3} \left
(\textstyle{\frac{1}{4}}B^g_{\mu\nu\alpha\beta} 
B^F_{\mu_1\nu_1,\alpha_1\beta_1}G^{\mu\mu_1;\nu\nu_1}
-           C^{RF}_{\mu\nu\rho;\alpha\beta}
C^{RF\mu\nu\rho}_{~~~~~~\alpha_1\beta_1}\right)
K^{\alpha\alpha_1;\beta_1\beta},
\end{equation}
up to subleading terms involving traces.
The complete bosonic invariant,
$
L_{4}  \equiv L_4^F+L_4^g+L_4^{Fg}+L_4^{FFFg},
$
is not necessarily in its  most unified form, but it
suggests some intriguing possibilities, especially in the matter 
sector. For example, it is worth noting that the ``C'' currents
can be unified into a unique current, which is the sum of the two, 
and their contributions to the invariant are simply its appropriate 
square.  The corresponding attempt for the BR sector unfortunately
does not quite work, at least with our choice of currents. We hope 
to return to this point elsewhere; instead 
we discuss some important  consequences of the very existence of 
this invariant, where elegance of its presentation is irrelevant.

Consider first the issue of renormalizability of D=11 supergravity. 
As we mentioned at the start, the work of \cite{BDetal} formally regarded  
as analytic continuation to D=11, states that the coefficient of 
a $2-$loop candidate counterterm is non-zero. Our result exhibits
this  invariant explicitly; taken together, they provide a compelling  
basis for the theory's nonrenormalizability.
In this connection a brief review of the divergence problem 
may be useful. For clarity, we choose to work in the framework of
dimensional  regularization, in which only logarithmic divergences
appear and consequently the local counterterm must have dimension zero
(including  dimensions of the coupling constants in the loop expansion). Now
a generic gravitational loop expansion proceeds in powers of
$\kappa^2$ (we will separately discuss the effect of the additional 
appearance of $\kappa$ in the CS vertex). At one loop, one would have 
$\triangle I_1\sim \kappa^0 \int dx^{11} \triangle L_1$; but there is no
candidate $\triangle L_1$ of dimension $11$, since odd dimension cannot 
be achieved  by a purely gravitational $\triangle L_1$, except at best
through a ``gravitational'' $\sim \epsilon\Gamma R R R
R$ or ``form-gravitational'' $\sim \epsilon A R R R R$ CS term \cite{Duff},
which would violate parity: Thus, if present, they would  represent an
anomaly, and so be finite anyway\footnote{In this connection we also note
that the presence of a Levi-Civita symbol $\epsilon$ usually does not
invalidate the use of dimensional regularization (or reduction)
schemes to the order we need. In any case our conclusions would also
apply, in  a more complicated way, in other regularization
schemes that preserve  SUSY.}. The  two-loop term would be $\triangle
L_2\sim\kappa^2\int d^{11}x \triangle L_2$, so that $\triangle L_2\sim 
[L]^{-20}$ which can be achieved 
(to lowest order in external lines)
by $\triangle L_2 \sim
\partial^{12} R^4$, where  $\partial^{12}$  means twelve explicit derivatives
spread among the 4 curvatures. There are no relevant  $2-$point 
$\sim \partial^{16} R^2 $ or  $3-$point $\sim \partial^{14} R^3 $
terms because the  
$R^2$ can be field-redefined away into the Einstein action in its 
leading part (to $h^2$ order, $E_4$ is a total divergence in any 
dimension!)
while $R^3$ cannot appear by SUSY. This latter fact was first
demonstrated in  D=4 but must therefore also apply in higher D
simply by  the brute force dimensional reduction argument. So the terms we
need
are, for their  4-graviton part, $L^g_4$ of (5) with twelve
explicit derivatives. The companions of $L_4^g$ in $L^{tot}_4$  will
simply appear with the same number of derivatives. It is easy to see
that the additional
$\partial^{12}$ can be inserted without spoiling
$SUSY$; indeed they appear as naturally as did 
multiplication by $s t u $ in  localizing the $M_4$ to $L_4$: for example,
$\partial^{12}$ might become, in momentum space 
language, $(s^6+t^6+u^6)$ or $(stu)^2$. This establishes the  structure of the 
$4-$point local counterterm candidate. As we
mentioned, its coefficient (more precisely that of $R^4$) is known and
non-vanishing at D=11 when calculated in the analytic continuation 
framework of  \cite{BDetal}, which is certainly correct through D=10.
Consider lastly possible invariants involving odd powers of $\kappa$ 
arising from the CS vertex. One might suppose that there is a class 
of $1-$loop
diagrams, consisting of a polygon (triangle or higher) with
form/graviton
segments and appropriate emerging external bosons at its vertices,
that could also have local divergences. The simplest  example would be
a form triangle  with three external $F-$lines $\sim\kappa^3\int d^{11} x 
\partial^9 \epsilon  A F F  $. 
This odd number of derivatives
cannot be
achieved and still yield a local scalar. 
 This argument also excludes
the one-loop polygon's  gravitational or form extensions such as $F^2
R$, $F R^2$ or even $F^3 R$ at this $\kappa^3$ level. 
One final comment:
nonrenormalizability had always been a reasonable guess as  
the fate of D=11 supergravity, given that it does not share the 
$N=4$  YM SUSY theory's
conformal invariance, because of the dimensional coupling constant 
$\kappa$. The opposite  guess, however, that some special 
(M$-$theory related?) property of this ``maximally maximal'' model
might keep it finite (at least to some higher order) could 
also have been reasonably entertained a priori, so this was an 
issue worth settling.

Perhaps more relevant to the future than the field theory's
ultraviolet behavior is the light that can be shed on ``nearby''
properties of M$-$theory, whatever its ultimate form. Given that 
D=11 supergravity is  its local limit, one would expect that there are 
local, ``zero-slope'' corrections that resemble  the corrections that 
D=10 string  theories make to their limiting D=10, supergravities.
Amongst other things, various brane effects might become
apparent in this way. Our local invariant (quite apart from the $\partial^n$
factors inserted for counterterm purposes) is then the simplest such
possible correction. As we saw, it shares with D=10 zero-slope limits
the same $t_8 t_8
R^4$ pure graviton term, but now acquires various
additional  form-dependent and spinorial contributions as well. 
A detailed version of our calculations will be  published
elsewhere. 

We are grateful to  Z. Bern and L. Dixon for very 
stimulating discussions about their and  our work, to M. Duff for 
a counterterm
conversation, to S. Fulling for useful information on invariant
bases and to J. Franklin for help with algebraic programming. This
work was supported by  NSF grant PHY-93-15811.

\end{document}